\documentclass[letterpaper,10pt,twocolumn]{article}

\usepackage{graphicx}
\usepackage{amsmath}
\usepackage{amssymb}

\newcommand{\dirac}{{\slash \negthinspace \negthinspace \negthinspace \nabla}}
\newcommand{\dd}{\textrm{d}}
\newcommand{\im}{{\mathbb{I}}{\mathrm{m}}}

\title{Quasinormal frequencies of the Dirac field in the massless topological black hole}

\author{A. L\'opez-Ortega\thanks{alopezo@ipn.mx} \\
Centro de Investigaci\'on en Ciencia Aplicada y Tecnolog\'{\i}a Avanzada. \\
	      Unidad Legaria. Instituto Polit\'ecnico Nacional. \\
              Calzada Legaria \# 694. Colonia Irrigaci\'on. Delegaci\'on Miguel Hidalgo. \\
	      M\'exico, D.\ F., M\'exico. \\
	      C.\ P.\  11500  
}

\begin{document}

\maketitle

\begin{abstract}

Motivated by the recent computations of the quasinormal frequencies of higher dimensional black holes we exactly calculate the quasinormal frequencies of the Dirac field propagating in $D$-dimensional ($D \geq 4$) massless topological black hole. From the exact values of the quasinormal frequencies for the fermion and boson fields we discuss whether the recently proposed bound on the relaxation time of a perturbed thermodynamical system is satisfied in $D$-dimensional massless topological black hole. Also we study the consequences of these results.\\


Motivados por el c\'alculo de las frecuencias cuasinormales de agujeros negros cuyo n\'umero de dimensiones $D$ es mayor o igual a cuatro, en el presente art\'iculo calculamos exactamente las frecuencias cuasinormales del campo de Dirac moviendose en el agujero negro topol\'ogico de masa cero con $D \geq 4$. Usando los valores exactos de las frecuencias cuasinormales para los fermiones y bosones discutimos si el l\'imite, recientemente propuesto, sobre el tiempo de relajamiento de un sistema termodin\'amico perturbado se satisface en el agujero negro topol\'ogico de masa cero con $D \geq 4$. Adicionalmente estudiamos algunas consecuencias de estos resultados.\\

\textit{Descriptores:} Modos cuasinormales, agujero negro topol\'ogico de masa cero, campo de Dirac, tiempo de relajamiento


\end{abstract}

\section{Introduction}
\label{section 1}

The physical systems for which we exactly solve their equations of motion can be expected to play a significant role in several research lines. For these physical systems we exactly calculate the physical quantities that for other systems we calculate by using approximate methods. Also in many research areas the physical insight that is obtained by studying the exactly solvable systems can be used to infer some details about the behavior of more complex physical systems.

The quasinormal modes (QNM) of a black hole are solutions to the equations of motion for a classical field that satisfy the appropriate radiation boundary conditions at the horizon and at the asymptotic region. The quasinormal frequencies (QNF) of a field are valuable quantities since these are determined by a few parameters of the black hole and the field \cite{Kokkotas:1999bd}--\cite{Nollert:1999bd}, for example, the QNF of the Kerr-Newman black hole are determined by the mass, angular momentum, and charge of the black hole and the mode of the field. Hence if we measure the QNF of a field then we can infer the values of the mass, angular momentum, and charge of the Kerr-Newman black hole. 

Also the QNM allow us to study the linear stability of the black holes, because if we find QNM whose amplitude increases in time, then the black hole may be unstable \cite{Kokkotas:1999bd}--\cite{Nollert:1999bd}. Recently the QNM have found applications in several research lines. For example, a) the AdS/CFT correspondence of string theory \cite{Berti:2009kk}, \cite{Horowitz:1999jd}, \cite{Birmingham:2001pj}, b) the determination of the area quantum of the black hole event horizon \cite{Hod:1998vk}, \cite{Maggiore:2007nq}, c) the expansion of functional determinants in some thermal spacetimes \cite{Denef:2009yy}, \cite{Denef:2009kn}, d) the expansion of the ``distant past'' Green functions used in self-force calculations \cite{Casals:2009zh}.

For many relevant spacetimes their QNF must be calculated approximately, hence we use numerical methods or perturbation methods \cite{Kokkotas:1999bd}--\cite{Nollert:1999bd}. Nevertheless, recently exact calculations of the QNF for several spacetimes have been presented. Among these we enumerate the following, a) three-dimensional static and rotating BTZ black holes \cite{Birmingham:2001pj}, \cite{Cardoso:2001hn}--\cite{Crisostomo:2004hj},  b) three-dimensional charged and rotating black holes of the Einstein-Maxwell-dilaton with cosmological constant theory \cite{Fernando:2003ai}--\cite{Fernando:2009tv}, c) two-dimensional dilatonic black hole \cite{Becar:2007hu}, \cite{LopezOrtega:2009zx}, d) five-dimensional dilatonic black hole \cite{Becar:2007hu}, \cite{LopezOrtega:2009zx}, e) $D$-dimensional de Sitter spacetime ($D \geq 3$) \cite{Du:2004jt}--\cite{LopezOrtega:2007sr}, f) BTZ black string \cite{Liu:2008ds}, g) Nariai spacetime \cite{Vanzo:2004fy}.\footnote{We notice that in Ref.\ \cite{Saavedra:2005ug} Saavedra presented an exact expression for the QNF of Unruh's acoustic black hole. The expression used in that reference for the effective metric of Unruh's acoustic black hole is valid near the horizon. For the asymptotic region of Unruh's acoustic black hole it is probable that we need to use a different approximation of the effective metric. Thus we believe that this problem deserves additional study. This issue was pointed out to the Author by the Referee.} In the following paragraphs we comment on another $D$-dimensional anti-de Sitter black hole for which the exact values of its QNF have been calculated.

We notice that the AdS/CFT correspondence of string theory motivated many studies on the QNF of anti-de Sitter black holes \cite{Berti:2009kk}, \cite{Horowitz:1999jd}, \cite{Birmingham:2001pj}, because this correspondence proposes that the QNF of the anti-de Sitter black holes determine the relaxation time of the dual conformal field theory \cite{Horowitz:1999jd}, \cite{Birmingham:2001pj}. See Ref.\ \cite{Birmingham:2001pj} for an explicit verification of this proposal in three-dimensional rotating BTZ black hole.

Furthermore we recall that in asymptotically anti-de Sitter spacetimes there are solutions of the Einstein equations that represent black holes whose horizons are negative curvature Einstein manifolds \cite{Vanzo:1997gw}--\cite{Lemos:1994xp}. These solutions are usually known as topological black holes and for some of these solutions the mass parameter can assume negative or zero values \cite{Vanzo:1997gw}--\cite{Lemos:1994xp}. 

Among these exact solutions of the Einstein equations there is one that has attracted a lot of attention. It is the asymptotically anti-de Sitter black hole whose mass is equal to zero \cite{Vanzo:1997gw}--\cite{Lemos:1994xp}. In the rest of the present paper we call it the massless topological black hole (MTBH). According to Ref.\ \cite{Birmingham:2006zx}, we can consider the MTBH as a higher dimensional generalization of the three-dimensional static BTZ black hole and we expect that it will play a significant role in future research. 

The metric of the $D$-dimensional MTBH is simple and as a consequence many of its physical properties can be calculated exactly \cite{Birmingham:2006zx}--\cite{Birmingham:2007yv}. For example, the QNF of the gravitational, Klein Gordon, and electromagnetic perturbations were calculated exactly in Refs.\ \cite{Birmingham:2006zx}, \cite{Aros:2002te}, and Sec.\ 6 of Ref.\ \cite{LopezOrtega:2007vu}, respectively. Also its stability against the three types of gravitational perturbations was proven in Refs.\ \cite{Gibbons:2002pq}, \cite{Birmingham:2007yv}. For numerical and analytical computations of the QNF for other topological black holes see Refs.\ \cite{Chan:1999sc}--\cite{Wang:2001tk}.

Here we exactly calculate the QNF of the Dirac field evolving in $D$-dimensional MTBH and thus we extend the results of Refs.\ \cite{Birmingham:2006zx}--\cite{LopezOrtega:2007vu}. The computation of the QNF for this fermion field is interesting because in some backgrounds the Dirac field behaves in a different way that the boson fields, for example, it is well known that in a rotating black hole the Dirac field does not show superradiant scattering \cite{Unruh:1974bw}--\cite{Iyer:1978du} in contrast to boson fields \cite{Chandrasekhar book}. Also notice that the QNF of the Dirac field allow us to discuss some additional details about the behavior of the MTBH under perturbations.

Note that in higher dimensional spacetimes, for the Dirac field we only know the QNF reported in Refs.\ \cite{LopezOrtega:2009zx}, \cite{LopezOrtega:2007sr}, \cite{Cho:2007zi}--\cite{Chakrabarti:2008xz}, thus for this fermion field its resonances have not been studied as extensively as for other fields. Hence this paper extends our knowledge on the QNM of the Dirac field in higher dimensional black holes.

This paper is organized as follows. In Sec.\ \ref{section 2} we find exact solutions to the Dirac equation in $D$-dimensional MTBH and using these solutions we exactly calculate the QNF of the Dirac field. Exploiting these results we enumerate some facts about the behavior of the MTBH under perturbations. In Sec.\ \ref{section 3} we investigate whether the fundamental QNF of the MTBH satisfy the bound recently proposed by Hod in Ref.\ \cite{Hod:2006jw}. In Sec.\ \ref{section 4}, following Chandrasekhar \cite{Chandrasekhar book}, in MTBH we write the Dirac equation as a pair of Schr\"odinger type differential equations and identify the effective potentials. Finally in Sec.\ \ref{section 5} we discuss the results obtained.

\section{QNF of the Dirac field}
\label{section 2}

The line element of a $G_{D-2}$-symmetric spacetime may be written as \cite{Kodama:2003jz}
\begin{equation} \label{eq: general metric}
\dd s^2 = F(r)^2 \dd t^2 -G(r)^2 \dd r^2 - H(r)^2 \dd \Sigma^2_{D-2},
\end{equation} 
where $F(r)$, $G(r)$, and $H(r)$ are functions only of the coordinate $r$ and $\dd \Sigma^2_{D-2}$ denotes the line element of a $(D-2)$-dimensional $G_{D-2}$-invariant base spacetime $\Sigma_{D-2}$, which depends only on the coordinates $\phi_i$, $i=1,2,\dots, D-2$.

Our aim is to calculate exactly the QNF of the Dirac field evolving in $D$-dimensional MTBH. Thus first we explicitly write the Dirac equation
\begin{equation} \label{eq: Dirac equation}
 i \dirac \psi = m \psi
\end{equation} 
in MTBH to find its exact solutions. Note that we follow the usual conventions, thus in formula (\ref{eq: Dirac equation}) the symbol $\dirac$ denotes the Dirac operator, $m$ stands for the mass of the Dirac field, and $\psi$ denotes the spinor of dimension $2^{[D/2]}$, where $[D/2]$ denotes the integer part of $D/2$ \cite{LopezOrtega:2009qc}, \cite{Gibbons:1993hg}--\cite{Cotaescu:1998ay}.

As is well known, in a $D$-dimensional $G_{D-2}$-symmetric spacetime with line element (\ref{eq: general metric}), the Dirac equation reduces to a pair of coupled partial differential equations in two variables (see for example Eqs.\ (30) of \cite{LopezOrtega:2009qc} and Refs.\ \cite{Gibbons:1993hg}--\cite{Cotaescu:1998ay} for more details)
\begin{align} \label{eq: Dirac equation general}
 \partial_t \psi_2 - \frac{F}{G} \partial_r \psi_2 & =  \left( i \kappa \frac{F}{H} -  i m F \right) \psi_1, \nonumber \\
\partial_t \psi_1 + \frac{F}{G} \partial_r \psi_1 & =  - \left( i \kappa \frac{F}{H} +  i m F \right) \psi_2,
\end{align} 
where $\kappa$ stands for the eigenvalues of the Dirac operator on the manifold $\Sigma_{D-2}$ with line element $\dd \Sigma^2_{D-2}$ and the functions $\psi_1$ and $\psi_2$ are the components of a two-dimensional spinor $\psi_{2D}$ which depends only on the coordinates $(t,r)$ of the $G_{D-2}$-symmetric spacetime with line element (\ref{eq: general metric}), that is
\begin{equation} \label{eq: spinor ansatz}
 \psi_{2D}(r,t)=\left( \begin{array}{c} \psi_1(r,t) \\ \psi_2(r,t) \end{array} \right).
\end{equation} 
We point out that in Eqs.\ (\ref{eq: Dirac equation general}) and in the rest of this paper we write the functions $\psi_1(r,t)$, $\psi_2(r,t)$, $F(r)$, $G(r)$, and $H(r)$ simply as $\psi_1$, $\psi_2$, $F$, $G$, and $H$, respectively. We shall use a similar convention for the functions to be defined in the rest of the present work.

The line element of the  $D$-dimensional MTBH is given by \cite{Vanzo:1997gw}--\cite{Lemos:1994xp}
\begin{equation} \label{eq: line element topological}
 \dd s^2 = \left(-1 + \frac{r^2}{L^2} \right) \dd t^2 - \frac{\dd r^2}{\left(-1 + \frac{r^2}{L^2} \right)} - r^2 \dd \Sigma_{D-2}^2,
\end{equation}  
where $r \in (L, + \infty)$, $L$ is related to the cosmological constant $\Lambda$ by
\begin{equation}
 L^2 = - \frac{(D-1)(D-2)}{2 \Lambda} ,
\end{equation}
and $\dd \Sigma^2_{D-2}$ stands for the line element of a $(D-2)$-dimensional compact space of negative curvature $\Sigma_{D-2}$ \cite{Vanzo:1997gw}--\cite{Lemos:1994xp}. Notice that the $(t,r)$ sector of the line element (\ref{eq: line element topological}) for the MTBH is similar to that of the three-dimensional static BTZ black hole with mass $M=1$. Taking into account this fact, it was proposed that the $D$-dimensional MTBH (\ref{eq: line element topological}) is a higher dimensional generalization of the three-dimensional static BTZ black hole \cite{Birmingham:2006zx}.

The QNM of the MTBH are solutions to the equations of motion for a field that are purely ingoing near the event horizon and since this black hole is asymptotically anti-de Sitter, we impose that at infinity the radial functions go to zero (Dirichlet's boundary condition) \cite{Birmingham:2006zx}, \cite{LopezOrtega:2007vu}. In this section we compute the QNF of the Dirac field propagating in $D$-dimensional MTBH to find out about the behavior of this black hole under fermion perturbations and compare with its behavior under boson perturbations. We note that the results of this section are an extension of those already published in Refs.\ \cite{Birmingham:2006zx}--\cite{LopezOrtega:2007vu}.

The line element of the $D$-dimensional MTBH (\ref{eq: line element topological}) has the same form that the line element of the $G_{D-2}$-symmetric spacetime (\ref{eq: general metric}). Thus making the appropriate identifications we get that the functions $F$, $G$, and $H$ for the MTBH are equal to
\begin{equation}                                                                                                      F=\frac{1}{G}=\left( -1 + \frac{r^2}{L^2}\right)^{1/2}, \qquad \qquad H=r.
\end{equation} 
Therefore in $D$-dimensional MTBH the coupled partial differential equations (\ref{eq: Dirac equation general}) reduce to  
\begin{align} \label{eq: Dirac equation topological}
\partial_t \psi_2 - \frac{z^2-1}{L} \partial_z \psi_2 &= (z^2-1)^{1/2} \left( \frac{i \kappa}{zL} - i m \right) \psi_1 , \nonumber \\ 
\partial_t \psi_1 + \frac{z^2-1}{L} \partial_z \psi_1 &= - (z^2-1)^{1/2} \left( \frac{i \kappa}{zL} + i m \right) \psi_2 ,
\end{align} 
where $z=r/L$ and therefore $z\in (1, +\infty)$. In what follows we write in detail the procedure used to solve exactly Eqs.\ (\ref{eq: Dirac equation topological}).

Choosing for the components $\psi_1$ and $\psi_2$ a harmonic time dependence of the form
\begin{align} \label{eq: psi 1 2 ansatz}
 \psi_1 (z,t) = \bar{R}_1(z)\, \textrm{e}^{-i \omega t } ,\nonumber \\
\psi_2 (z,t) = R_2(z)\, \textrm{e}^{-i \omega t },
\end{align} 
and defining $\tilde{\omega} = \omega L$, $\tilde{m} = m L$, and $K=- i \kappa$ we get that the system of partial differential equations (\ref{eq: Dirac equation topological}) transforms into the  coupled system of ordinary differential equations for the functions $R_2$ and $ R_1 = -i \bar{R}_1 $  
\begin{align} \label{eq: radial two topological}
(z^2 - 1)\frac{\dd R_2}{\dd z} + i \tilde{\omega} R_2 & = (z^2 -1)^{1/2} \left( \frac{iK}{z} -  \tilde{m} \right) R_1,   \\ 
(z^2 - 1)\frac{\dd R_1}{\dd z} - i \tilde{\omega} R_1 & = -(z^2 -1)^{1/2} \left( \frac{iK}{z} + \tilde{m} \right) R_2. \nonumber
\end{align}
 
If we make the following ansatz for the functions $R_1$ and $R_2$ (see formulas (26) of Ref.\ \cite{LopezOrtega:2007sr} for a similar ansatz for the radial functions of the Dirac field evolving in $D$-dimensional de Sitter spacetime)
\begin{align}
 R_1 (z)= (z^2 -1)^{-1/4} (z+1)^{1/2} \tilde{R}_1(z), \nonumber \\
R_2 (z) = (z^2 -1)^{-1/4} (z-1)^{1/2} \tilde{R}_2(z),
\end{align}
then we find that the functions $\tilde{R_1}$ and $\tilde{R_2}$ satisfy
\begin{align}
 (z^2-1) \frac{\dd \tilde{R}_2}{\dd z} + \left( i \tilde{\omega} + \tfrac{1}{2}\right) \tilde{R}_2 &= \left( \frac{i K}{z} - \tilde{m} \right)(z+1) \tilde{R}_1,  \nonumber \\
(z^2-1) \frac{\dd \tilde{R}_1}{\dd z} - \left( i \tilde{\omega} + \tfrac{1}{2}\right) \tilde{R}_1 &= - \left( \frac{i K}{z} + \tilde{m} \right)(z-1) \tilde{R}_2. 
\end{align}

Next, we define the functions $f_1$ and $f_2$ by
\begin{equation}  \label{eq: f1 f2 definition}
 f_1 (z)= \tilde{R}_1 (z)+ \tilde{R}_2(z), \quad f_2 (z)= \tilde{R}_1(z) - \tilde{R}_2(z),
\end{equation} 
to obtain that these functions must be solutions to the coupled system of ordinary differential equations 
\begin{align}  \label{eq: coupled topological f1 f2}
(z^2 -1) \frac{\dd f_1}{\dd z} + \left(\tilde{m} z - \frac{i K}{z} \right) f_1 = \left( i \tilde{\omega} + \tfrac{1}{2} + i K - \tilde{m} \right) f_2,  \nonumber \\
(z^2 -1) \frac{\dd f_2 }{\dd z} - \left(\tilde{m} z - \frac{i K}{z} \right) f_2 = \left( i \tilde{\omega} + \tfrac{1}{2} - i K + \tilde{m} \right) f_1 .
\end{align}
From Eqs.\ (\ref{eq: coupled topological f1 f2}) we obtain that the functions $f_1$ and $f_2$ satisfy the decoupled ordinary differential equations 
\begin{align} \label{eq: Dirac topological f1 f2}
(z^2 -1)^2 \frac{\dd^2 f_1}{\dd z^2} &+ 2z(z^2-1)\frac{\dd f_1}{\dd z} + (z^2-1)\left(\tilde{m} + \frac{iK}{z^2} \right) f_1 \nonumber \\
&- \left( \tilde{m}^2 z^2 - 2 \tilde{m} i K - \frac{K^2}{z^2} \right)  f_1 \nonumber \\
& = \left( \left( i \tilde{\omega} + \tfrac{1}{2} \right)^2 - (i K - \tilde{m})^2  \right) f_1,  \nonumber \\
(z^2 -1)^2 \frac{\dd^2 f_2}{\dd z^2} &+ 2z(z^2-1)\frac{\dd f_2}{\dd z} - (z^2-1)\left(\tilde{m} + \frac{iK}{z^2} \right) f_2 \nonumber \\
&- \left( \tilde{m}^2 z^2 - 2 \tilde{m} i K - \frac{K^2}{z^2} \right) f_2 \nonumber \\
& = \left( \left( i \tilde{\omega} + \tfrac{1}{2} \right)^2 - (i K - \tilde{m})^2  \right) f_2 .  
\end{align}

To solve Eqs.\ (\ref{eq: Dirac topological f1 f2}) we make the changes of variables $x=z^2$ and $u=(x-1)/x$, and take the functions $f_1$ and $f_2$ in the form
\begin{align}
 f_1(u)&= u^{B_1} (1-u)^{F_1} \hat{R}_1(u), \nonumber \\ 
f_2(u)&= u^{B_2} (1-u)^{F_2} \hat{R}_2(u) ,
\end{align}
where 
\begin{align}
B_1 =& B_2 =  \left\{ \begin{array}{l} \frac{i \tilde{\omega}}{2} + \frac{1}{4} , \\ \\ - \frac{i \tilde{\omega}}{2} - \frac{1}{4} , \end{array}\right.  \\ \nonumber
 F_1 = & \left\{ \begin{array}{l}  \frac{1}{4} + \frac{1}{2} \sqrt{\tilde{m}^2 - \tilde{m} + \tfrac{1}{4} } ,  \nonumber \\ \\  
\frac{1}{4} - \frac{1}{2} \sqrt{\tilde{m}^2 - \tilde{m} + \tfrac{1}{4}} , \end{array} \right.   \\ \nonumber 
 F_2 = & \left\{ \begin{array}{l} \frac{1}{4} + \frac{1}{2} \sqrt{\tilde{m}^2 + \tilde{m} + \tfrac{1}{4} } ,  \\ \\ 
 \frac{1}{4} - \frac{1}{2} \sqrt{\tilde{m}^2 + \tilde{m} +  \tfrac{1}{4}} , \end{array} \right.  \nonumber
\end{align}
to find that the functions $\hat{R}_1$ and $\hat{R}_2$ must be solutions of the hypergeometric differential equation \cite{b: Abramowitz book}, \cite{b: Wang book} 
\begin{equation} \label{e: hypergeometric differential equation}
u(1-u) \frac{\dd^2 f}{\dd u^2} + (c - (a +b + 1)u)\frac{\dd f}{{\rm d}u} - a b f   = 0.
\end{equation} 
If the parameter $c$ is not an integer, then the solutions of Eq.\ (\ref{e: hypergeometric differential equation}) are given in terms of the standard hypergeometric functions ${}_{2}F_{1}(a,b;c;u)$ \cite{b: Abramowitz book}, \cite{b: Wang book}. 

For the functions $\hat{R}_1$ and $\hat{R}_2$ the quantities $a$, $b$, and $c$ of Eq.\ (\ref{e: hypergeometric differential equation}) are equal to ($a_i$, $b_i$, and $c_i$ correspond to the function $\hat{R}_i$, $i=1,2$)
\begin{align} \label{eq: constants hypergeometric f1 f2}
a_1 & = B_1 + C_1 + \tfrac{1}{4} +  \tfrac{1}{2} \sqrt{\tilde{m}^2 - \tilde{m} + \tfrac{1}{4}}, \nonumber \\
b_1 & = B_1 - C_1 + \tfrac{3}{4} +  \tfrac{1}{2} \sqrt{\tilde{m}^2 - \tilde{m} + \tfrac{1}{4} }, \nonumber \\
c_1 & = 2 B_1 +1 ,  \\
a_2 & = B_2 + C_2 + \tfrac{1}{4} +  \tfrac{1}{2} \sqrt{\tilde{m}^2 + \tilde{m} + \tfrac{1}{4} }, \nonumber \\
b_2 & = B_2 - C_2 + \tfrac{3}{4} +  \tfrac{1}{2} \sqrt{\tilde{m}^2 + \tilde{m} + \tfrac{1}{4}}, \nonumber \\
c_2 & = 2 B_2 +1  , \nonumber
\end{align} 
where the quantities $C_1$ and $C_2$ take the values
\begin{align} \label{eq: constants topological}
 C_1 = & \left\{ \begin{array}{l} \frac{1}{2} + \frac{i K}{2} , \\ \\ - \frac{i K}{2} , \end{array} \right.  \qquad \quad
C_2 =  \left\{ \begin{array}{l} \frac{i K}{2} , \\ \\ \frac{1}{2} - \frac{i K}{2} . \end{array}\right. 
\end{align}

At this point we notice that the coordinate $x$ lies in the range $x \in (1,+ \infty)$. Hence the variable $u$ satisfies $u \in (0,1)$. Also the tortoise coordinate of the MTBH is \cite{LopezOrtega:2007vu}
\begin{equation} \label{eq. tortoise coordinate}
 r_* = \int \left(-1 + \frac{r^2}{L^2} \right)^{-1} \dd r = - L \,\textrm{arccoth} (z),
\end{equation} 
thus $r_* \in (-\infty, 0)$, $r_* \to -\infty$ near the event horizon and $r_* \to 0$ near infinity. From these definitions of the coordinates $u$ and $r_*$ we get that
\begin{align} \label{eq: topological relations coordinates}
 \textrm{as } \quad & r_* \to - \infty,  & u \approx & \,  \textrm{e}^{2 r_*/L}, \qquad  \textrm{and} \\
\textrm{as } \quad & r_* \to 0,  & u \approx & \, 1. \qquad  \qquad \qquad \quad \,\,\,  \nonumber
\end{align}

Now we use these results to compute the QNF of the Dirac field exactly. First let us study the function $f_1$. We choose the quantities $C_1$, $B_1$, and $F_1$ as $C_1=1/2 + iK/2$, $B_1 = i\tilde{\omega}/2 + 1/4$, and $F_1 = 1/4 + \sqrt{\tilde{m}^2 - \tilde{m} + 1/4}/2$. If we assume that the quantity $c_1$ is not an integer, then we obtain that the function $f_1$ is equal to
\begin{align}
 &f_1  = (1-u)^{F_1} \left\{  \mathbb{D}_1 u^{i \tilde{\omega}/2 +1/4} {}_{2}F_{1}(a_1,b_1;c_1;u)  \right.  \\
& \left. + \, \mathbb{E}_1 u^{- i \tilde{\omega}/2 - 1/4} {}_{2}F_{1}(a_1-c_1+1,b_1-c_1+1;2-c_1;u) \right\} , \nonumber
\end{align}
where $\mathbb{D}_1$ and $\mathbb{E}_1$ are constants. Taking into account expressions (\ref{eq: topological relations coordinates}) we find that near the horizon the function $f_1$ behaves as
\begin{equation}
 f_1 \approx  \mathbb{D}_1 \textrm{e}^{i \omega r_* + r_*/(2L)} + \mathbb{E}_1 \textrm{e}^{-i \omega r_* - r_*/(2L)};
\end{equation} 
thus in order to have a purely ingoing wave near the event horizon we must impose the condition $\mathbb{D}_1=0$ \cite{Birmingham:2006zx}--\cite{LopezOrtega:2007vu}. Hence the function $f_1$ becomes
\begin{align} \label{eq: f1 1-z topological}
 f_1  &=  \mathbb{E}_1 u^{- i \tilde{\omega}/2 - 1/4} (1-u)^{1/4 + \sqrt{\tilde{m}^2 - \tilde{m} + 1/4}/2} \nonumber \\    
& \times {}_{2}F_{1}(a_1-c_1+1,b_1-c_1+1;2-c_1;u) \\  
 &     =  \mathbb{E}_1 u^{- i \tilde{\omega}/2 - 1/4}  (1-u)^{1/4 + \sqrt{\tilde{m}^2 - \tilde{m} + 1/4}/2} \nonumber \\ 
& \times {}_{2}F_{1}(\alpha_1,\beta_1;\gamma_1;u). \nonumber 
\end{align}

We recall that if the quantity $c-a-b$ is not an integer then the hypergeometric function ${}_{2}F_{1}(a,b;c;u)$ satisfies \cite{b: Abramowitz book}, \cite{b: Wang book} 
\begin{align} \label{e: hypergeometric property z 1-z}
 {}_2F_1(a,b;c;u)  &= \frac{\Gamma(c) \Gamma(c-a-b)}{\Gamma(c-a) \Gamma(c - b)}  \nonumber \\
& \times {}_2 F_1 (a,b;a+b+1-c;1-u) \nonumber \\
&+ \frac{\Gamma(c) \Gamma( a + b - c)}{\Gamma(a) \Gamma(b)} (1-u)^{c-a -b} \\
&\times {}_2F_1(c-a, c-b; c + 1 -a-b; 1 -u),\nonumber
\end{align} 
where $\Gamma(x)$ stands for the gamma function. Hence if the quantity $\gamma_1 - \alpha_1 - \beta_1$ is not an integer, then we write the function $f_1$ of formula (\ref{eq: f1 1-z topological}) as
\begin{align} \label{eq: topological f1 1-u}
  f_1  &= \mathbb{E}_1 u^{- i \tilde{\omega}/2 - 1/4} \left[ \frac{\Gamma(\gamma_1) \Gamma(\gamma_1 - \alpha_1 - \beta_1)}{\Gamma(\gamma_1 - \alpha_1) \Gamma(\gamma_1 - \beta_1)}  \right.  \\ 
& \left. \times (1-u)^{ 1/4 +  \sqrt{\tilde{m}^2 - \tilde{m} + 1/4 }/2} \right. \nonumber \\
& \left. \times {}_2 F_1 (\alpha_1,\beta_1;\alpha_1+\beta_1+1-\gamma_1;1-u) \right. \nonumber \\  
& +  \frac{\Gamma(\gamma_1) \Gamma( \alpha_1 + \beta_1 - \gamma_1)}{\Gamma(\alpha_1) \Gamma(\beta_1)} 
  (1-u)^{1/4 - \sqrt{\tilde{m}^2 - \tilde{m} + 1/4}/2} \nonumber \\ 
&\left. \times {}_2F_1(\gamma_1-\alpha_1, \gamma_1-\beta_1; \gamma_1 + 1 -\alpha_1 - \beta_1; 1 -u) \right]. \nonumber 
\end{align}

Due to the MTBH being asymptotically anti-de Sitter the QNM boundary conditions at infinity demands that $f_1 \to 0$ as $u \to 1$ \cite{Birmingham:2006zx}, \cite{LopezOrtega:2007vu}. From the expression (\ref{eq: topological f1 1-u}) we note that the first term in square brackets vanishes as $u \to 1$. The second term vanishes for $1/2 > \sqrt{\tilde{m}^2 - \tilde{m} + 1/4}$, (that is, for $1>\tilde{m}$). Thus the function $f_1$ goes to zero as $u \to 1$, and therefore if $1>\tilde{m}$ the boundary condition at infinity does not impose any restriction on the frequencies, that is, there is a continuum of frequencies that satisfy the boundary condition at infinity of the QNM. For $\tilde{m} \geq 1$, in order that $f_1 \to 0$ as $u \to 1$ we must impose the condition
\begin{equation} \label{eq: conditions topological alpha}
 \alpha_1 = -n_1, \quad \textrm{or} \quad \beta_1 = -n_1, \qquad n_1=0,1,2,\dots
\end{equation} 
Therefore for $\tilde{m} \geq 1$, from Eqs.\ (\ref{eq: conditions topological alpha}) we find that the QNF of the function $f_1$ are equal to
\begin{align} \label{eq: QN frequencies f1}
 \tilde{\omega}_1 & = K - i \left( 2 n_1 + 1 + \sqrt{\tilde{m}^2 - \tilde{m} + \tfrac{1}{4}} \right), \qquad \textrm{or} \nonumber \\
 \tilde{\omega}_1 & = - K - i \left( 2 n_1 + \sqrt{\tilde{m}^2 - \tilde{m} +  \tfrac{1}{4} } \right) ,
\end{align} 
whereas for $1>\tilde{m}$ there is a continuum of QNF. 

To calculate the QNF of the function $f_2$ we choose the quantities $C_2$, $B_2$, and $F_2$ as  $C_2= iK/2$, $B_2 = i\tilde{\omega}/2 + 1/4$, and $F_2 = 1/4 +  \sqrt{\tilde{m}^2 + \tilde{m} + 1/4}/2$. A similar method to that used for the function $f_1$ allows us to find that for all $\tilde{m}$ the QNF of the function $f_2$ are ($ n_2=0,1,2,\dots$)
\begin{align} \label{eq: QN frequencies f2}
 \tilde{\omega}_2 & = K - i \left( 2 n_2 + \sqrt{\tilde{m}^2 + \tilde{m} +  \tfrac{1}{4} }  \right), \quad  \quad \textrm{or} \nonumber \\
 \tilde{\omega}_2 & = - K - i \left( 2 n_2 + 1 + \sqrt{\tilde{m}^2 + \tilde{m} +  \tfrac{1}{4} } \right). 
\end{align} 

From formulas (\ref{eq: f1 f2 definition}) we find that the functions $\hat{R}_1$ and $\hat{R}_2$ are linear combinations of the functions $f_1$ and $f_2$, therefore only the QNF that are equal for both functions $f_1$ and $f_2$ will be QNF of the Dirac field in $D$-dimensional MTBH. Thus when $\tilde{m} < 1$, for the function $f_1$ we find a continuum of QNF, but for the function $f_2$ we only find the QNF (\ref{eq: QN frequencies f2}). Hence for $\tilde{m} < 1$ the QNF of the Dirac field are equal to
\begin{align} \label{eq: QN frequencies topological} 
 \omega& =  \frac{K}{L} - \frac{i}{L} \left(2 n + \frac{1}{2} +\tilde{m} \right), \qquad  n=0,1,2,\dots \nonumber \\
 \omega& = -\frac{K}{L} - \frac{i }{L} \left(2 n  +\frac{3}{2} +\tilde{m} \right) .
\end{align} 

When $\tilde{m} \geq 1$ for the function $f_1$ we find the QNF (\ref{eq: QN frequencies f1}), whereas for the function $f_2$ we find the QNF (\ref{eq: QN frequencies f2}). After some simplifications we find that for $\tilde{m} \geq 1$ the QNF frequencies of the Dirac field are also determined by the expressions (\ref{eq: QN frequencies topological}). Thus in MTBH formulas (\ref{eq: QN frequencies topological}) give the QNF of the Dirac field for any value of the mass $\tilde{m}$.  In the massless limit the QNF (\ref{eq: QN frequencies topological}) reduce to
\begin{align} \label{eq: QN frequencies topological massless}
 \omega & =  \frac{K}{L} - \frac{i}{L} \left(2 n + \frac{1}{2} \right),  \nonumber \\
 \omega & = - \frac{K}{L} - \frac{i}{L} \left(2 n  +\frac{3}{2} \right) .
\end{align} 

For QNF (\ref{eq: QN frequencies topological}) and (\ref{eq: QN frequencies topological massless}) we find that $ \im \,( \tilde{\omega} ) < 0$, hence these QNM decay in time. Thus the $D$-dimensional MTBH is linearly stable against Dirac perturbations. Something similar happens for the QNF of the electromagnetic and gravitational perturbations \cite{Birmingham:2006zx}, \cite{LopezOrtega:2007vu}. The stability of the MTBH against the gravitational perturbations was shown in Refs.\ \cite{Gibbons:2002pq}, \cite{Birmingham:2007yv}.

As we previously commented, in Refs.\ \cite{Birmingham:2006zx}--\cite{LopezOrtega:2007vu} it was calculated the QNF of the  gravitational, electromagnetic, and minimally coupled massless Klein-Gordon perturbations. The values obtained for the QNF of these fields are 
\begin{equation} \label{eq: massless bosons}
 \omega =\pm \frac{\xi}{L} - \frac{2i}{L}\left( n + \frac{\mathbb{A}}{4} \right) ,
\end{equation} 
where the quantity $\mathbb{A}$ takes on the values
\begin{align} \label{eq: A values}
 \mathbb{A}=  & \left\{ \begin{array}{l}  D-1 \, \, \, \, \textrm{for the vector type gravitational } \\ 
\quad \qquad \,\, \, \, \,  \textrm{and electromagnetic perturbations}, \\
 |D-5|+2  \, \, \, \, \textrm{for the scalar type gravitational } \\
\quad \qquad \,\, \, \, \,  \textrm{and electromagnetic perturbations}, \\
 D+1  \, \, \textrm{for the tensor type gravitational } \\
\quad \quad \,\, \, \, \textrm{perturbation ($D\geq5$) and minimally } \\
\quad \quad \,\, \, \, \textrm{ coupled massless Klein-Gordon field,}
 \end{array} \right.
\end{align}   
the quantity $\xi$ depend on the perturbation type and is related to the eigenvalues of the Laplacian on the manifold $\Sigma_{D-2}$ \cite{Birmingham:2006zx}--\cite{LopezOrtega:2007vu}.  

For the non-minimal coupled to gravity massive Klein-Gordon field the QNF are equal to \cite{Aros:2002te}
\begin{equation} \label{eq: Klein Gordon massive}
 \omega = \pm \frac{\xi}{L} - \frac{i}{L}\left(2n + 1  + \sqrt{\left(\tfrac{D-1}{2}\right)^2 + m_{eff}^2 L^2} \right),
\end{equation} 
where $m_{eff}^2 = m^2 - \gamma D(D-2)/(4 L^2)$, $m$ denotes the mass of the Klein-Gordon field, and $\gamma$ is the coupling constant between the scalar curvature and the Klein-Gordon field. Notice that in Ref.\ \cite{Aros:2002te} it was chosen a different time parameter to that used in the present paper. This fact implies that the QNF (\ref{eq: Klein Gordon massive}) have an additional factor of $1/L$ to the QNF reported in Ref.\ \cite{Aros:2002te}.

From formulas (\ref{eq: QN frequencies topological})--(\ref{eq: Klein Gordon massive}) we find that for the Dirac field the imaginary part of the QNF (\ref{eq: QN frequencies topological}) and (\ref{eq: QN frequencies topological massless}) does not depend on the spacetime dimension, unlike boson fields the imaginary part of their QNF (\ref{eq: massless bosons}) and (\ref{eq: Klein Gordon massive}) shows an explicit dependence on the spacetime dimension. Thus for the Dirac field the decay time $\tau_d = 1/|\im (\omega)|$ depends on the mode number $n$ and it does not depend on the spacetime dimension. In contrast for the boson fields the decay time is inversely proportional to the spacetime dimension, thus for a given boson field and fixed mode number, the decay time decreases as the spacetime dimension increases. Hence in $D$-dimensional MTBH the decay time for the Dirac field and the decay time for the boson fields show a different behavior when the spacetime dimension changes.

Furthermore, from formulas (\ref{eq: massless bosons}) and (\ref{eq: A values}) for the massless boson fields with mode number fixed, and for $D \geq 5$  we find that the tensor type gravitational perturbation and minimally coupled massless Klein-Gordon field decay faster than vector type and scalar type electromagnetic and gravitational perturbations.

From QNF (\ref{eq: QN frequencies topological massless}) and (\ref{eq: massless bosons}) we see that for $D \geq 6$ the decay time of the massless Dirac field is greater than the decay time of the massless boson fields. Thus for $D \geq 6$ the massless boson fields decay faster than massless Dirac field. Also for $D=5$ the tensor type gravitational perturbation and minimally coupled massless Klein-Gordon field decay faster than the other massless boson fields and massless Dirac field. For $D=4$ we find that the minimally coupled massless Klein-Gordon field decays faster than the electromagnetic, gravitational, and massless Dirac perturbations.

It is convenient to note that for the massive Klein-Gordon and Dirac fields the imaginary part of the QNF depends on the mass of the field. Taking into account formulas (\ref{eq: QN frequencies topological}) and (\ref{eq: Klein Gordon massive}) we find that if the mass of the Dirac and the minimally coupled Klein-Gordon fields are equal and the condition
\begin{equation}
 m L < \left(\frac{D}{2} - 1 \right)\frac{D}{2}
\end{equation} 
is satisfied, then the minimally coupled Klein-Gordon field decays faster than the Dirac field. 

In MTBH the oscillation frequencies of the boson and fermion fields do not depend on the mass of the field. For the boson fields the oscillation frequencies are determined by the eigenvalues of the Laplace operator on the negative curvature manifold $\Sigma_{D-2}$, whereas for the Dirac field the oscillation frequencies are determined by the eigenvalues of the Dirac operator on $\Sigma_{D-2}$. 

Thus for a complete determination of the QNF (\ref{eq: QN frequencies topological}) for the Dirac field moving in MTBH, we need to know the eigenvalues of the Dirac operator on the base manifold $\Sigma_{D-2}$ with metric $\dd \Sigma_{D-2}^2$. We expect that the event horizon of a black hole will be a compact and orientable manifold \cite{Aros:2002rk}. For the MTBH the negative curvature manifold $\Sigma_{D-2}$ usually is a quotient of the form $H^{D-2}/G$, where $G$ is a freely acting discrete subgroup of the isometry group for the $(D-2)$-dimensional hyperbolic space $H^{D-2}$. Therefore for the QNF (\ref{eq: QN frequencies topological}) of the MTBH we need to find the spectrum of the Dirac operator on a compact spin manifold of hyperbolic type. Regarding the spectrum of the Dirac operator on hyperbolic manifolds we know the following facts.

In contrast to the Laplace operator, the spectrum of the Dirac operator depends on the geometry of the manifold and the spin structure, which is a topological object that we need to define spinors \cite{Bar:2000}, \cite{b: Friedrich book}. In general, the spin structure of a spin manifold is not unique, for example the circle $S^1$ has two spin structures, but note that some manifolds do not admit even a spin structure, for example the complex projective plane $\mathbb{C P}^2$ \cite{Bar:2000}, \cite{b: Friedrich book}.

The hyperbolic space $H$ has an unique spin structure (it is due to that the hyperbolic space is contractible) \cite{Camporesi:2001}. It is known that on the hyperbolic space for the Dirac operator the discrete spectrum is empty and its continuous spectrum is $\mathbb{R}$ \cite{Bar:2000}, \cite{Camporesi:2001}. We note that the conventions used in Refs.\ \cite{Bar:2000}, \cite{Camporesi:2001} and the present paper are different. In the conventions that we use here the eigenvalues of the Dirac operator on the hyperbolic space are purely imaginary as in Ref.\ \cite{Camporesi:1995fb} (and therefore $K \in \mathbb{R}$), whereas in Refs.\ \cite{Bar:2000}, \cite{Camporesi:2001} the eigenvalues of the Dirac operator on the hyperbolic space are real numbers.

If the manifold is compact, then general elliptic theory asserts that the spectrum of the Dirac operator is discrete \cite{Bar:2000}, \cite{b: Friedrich book}. Thus we expect that on the base manifold $\Sigma_{D-2}$ of the MTBH the eigenvalues of the Dirac operator are discrete. Furthermore for a $D$-dimensional compact manifold $\Sigma$ the eigenvalues $\kappa$ of the Dirac operator satisfy the Weyl asymptotic law \cite{Bar:2000}
\begin{equation}
 \lim_{\kappa \to \infty} \frac{N(\kappa)}{\kappa^D} = \frac{2^{[D/2]} \textrm{vol}(\Sigma)}{(4 \pi)^{D/2} \Gamma\left(\frac{D}{2} +1\right)},
\end{equation} 
where $\textrm{vol}(\Sigma)$ is the volume of the $D$-dimensional manifold $\Sigma$ and $N(\kappa)$ is the number of eigenvalues whose modulus is $\leq \kappa$. 

On a compact symmetric manifold with a homogeneous spin structure the square of the Dirac operator $\dirac^2$ satisfies \cite{b: Friedrich book}, \cite{b: Ginoux book}
\begin{equation}
 \dirac^2 = \Omega + \frac{\mathcal{R}}{8},
\end{equation} 
where $\Omega$ is the Casimir operator of the isometry group and $\mathcal{R}$ is the scalar curvature of the compact symmetric manifold. Therefore for these manifolds the computation of the spectrum for the square of the Dirac operator can be done by algebraic methods. Also on these manifolds the spectrum of the Dirac operator is symmetric with respect to the origin and the spectrum of the Dirac operator is determined by the spectrum of its square. Nevertheless there are technical difficulties and the spectrum of the Dirac operator is explicitly known for a small number of manifolds \cite{b: Ginoux book}.

As far as we know for compact hyperbolic manifolds the spectrum of the Dirac operator is calculated exactly for the manifold $\Sigma = PSL_2(\mathbb{R})/\Gamma$, where $PSL_2(\mathbb{R})$ is the projective special linear group of $\mathbb{R}^2$ and $\Gamma$ is a co-compact Fuchsian subgroup \cite{b: Ginoux book}, \cite{Seade:1987}. The complicated spectrum of the Dirac operator on $\Sigma = PSL_2(\mathbb{R})/\Gamma$ appears in Theorem 2.2.3 of Ref.\ \cite{b: Ginoux book}. Notice that the case relevant to our work is when the parameter $t$ of Theorem 2.2.3 is equal to $1$ and therefore the manifold $PSL_2(\mathbb{R})/\Gamma$ has negative constant sectional curvature.\footnote{For the related case of the so called plane symmetric black hole, it is convenient to notice that the spectrum of the Dirac operator on the higher dimensional flat tori has been calculated (see Theorem 4.1 of Ref.\ \cite{Bar:2000} and Theorem 2.1.1 of Ref.\ \cite{b: Ginoux book}). We point out that the flat tori admits several spin structures and the spectrum of the Dirac operator depends on the spin structure \cite{b: Ginoux book}. For other examples of flat manifolds for which the spectrum of the Dirac operator is calculated exactly see Chapter 2 of Ref.\ \cite{b: Ginoux book}.}

We notice that for the Dirac operator eigenvalue estimates can be found in several manifolds for which an exact calculation of the spectrum is not possible  \cite{b: Ginoux book}. We believe that the following result is relevant for our work.

In Proposition 2 of Ref.\ \cite{Baum:1991} it is asserted that for a compact and oriented two dimensional surface $\Sigma$ of genus $g \neq 1$ there is an eigenvalue $\kappa$ of the Dirac operator that satisfies 
\begin{equation}
 \mid {\kappa} \mid \leq c(g) \, \textrm{max} \{ \textrm{principal curvatures of} \,\, \Sigma \},
\end{equation} 
where
\begin{align}
c(g) = & \left\{ \begin{array}{l} 1 \, \, \, \, \, \textrm{if}\, \, \, \, \, g = 0, \\ 
3 \,\, \, \, \, \textrm{if}\, \, \, \, \, g = 2,3, \\
4 \,\, \, \, \, \textrm{if} \,\, \,\, \,  g \geq 4 . \end{array} \right.
\end{align}   
For $g \geq 2$ this result is pertinent for the four-dimensional MTBH. We do not know similar estimates for the eigenvalues of the Dirac operator on higher dimensional compact hyperbolic manifolds. 

From these comments it is deduced that in the mathematical literature we do not find many calculations on the eigenvalues of the Dirac operator on compact hyperbolic manifolds and we believe that the computation of these quantities is a challenging mathematical problem.

In Ref.\ \cite{Birmingham:2001pj} it was shown that the momentum space poles of the retarded correlation functions in the dual conformal field theory and the QNF of the three-dimensional BTZ black hole are identical. Calculating whether something similar happens for the QNF of the $D$-dimensional MTBH is an interesting problem.

\section{Hod's bound}
\label{section 3}

Taking into account quantum information theory and thermodynamic concepts, in Ref.\ \cite{Hod:2006jw} Hod found a bound on the relaxation time $\tau$ of a perturbed thermodynamic system. This bound is
\begin{equation} \label{eq: Hod bound one}
 \tau \geq \tau_{min} = \frac{\hbar}{\pi T},
\end{equation} 
where $\tau_{min}$ stands for the minimum relaxation time and $T$ denotes the temperature of the thermodynamic system. This bound is called ``TTT bound'' (time times temperature bound) by Hod in Ref.\ \cite{Hod:2006jw}.

In Ref.\ \cite{Hod:2006jw} it was shown that strong self-gravity systems, as the black holes, are the appropriate systems to test the TTT bound (\ref{eq: Hod bound one}). For a black hole the TTT bound states that at least for the fundamental QNF the following inequality is satisfied \cite{Hod:2006jw}
\begin{equation} \label{eq: Hod's inequality}
 \frac{\hbar \omega_I}{\pi T_{H}}  \leq 1,
\end{equation} 
where $\omega_I$ is the absolute value of the imaginary part of the fundamental QNF and $T_{H}$ is Hawking's temperature of the black hole (see Refs.\ \cite{Hod:2006jw}, \cite{Hod:2007tb}--\cite{Hod:2008zz} for more details). The fundamental QNM is the least damped mode of the black hole and it determines its relaxation time scale \cite{Kokkotas:1999bd}--\cite{Nollert:1999bd}. 

The Hawking temperature of the MTBH is equal to \cite{Vanzo:1997gw}--\cite{Lemos:1994xp}
\begin{equation} 
 T_H = \frac{\hbar}{2 \pi L},
\end{equation} 
and from the QNF (\ref{eq: QN frequencies topological massless}) of the massless Dirac field we find
\begin{equation} \label{eq: Dirac field ratio}
 \frac{\hbar \omega_I}{\pi T_H} = 1 \qquad \quad \textrm{and} \qquad \quad  \frac{\hbar \omega_I}{\pi T_H} = 3.
\end{equation} 
We see that the first expression in formulas (\ref{eq: Dirac field ratio}) saturates the inequality (\ref{eq: Hod's inequality}) and that the second expression does not satisfy the previously mentioned inequality.

Furthermore, from QNF (\ref{eq: massless bosons}) of the massless boson fields we obtain that
\begin{equation} \label{eq: bound bosons}
 \frac{\hbar \omega_I}{\pi T_H} = \mathbb{A}.
\end{equation} 
Hence, taking into account the values of the quantity $ \mathbb{A}$ given in formula (\ref{eq: A values}), for $D \geq 4$ we find that in MTBH the fundamental QNF of the massless bosons do not satisfy the inequality (\ref{eq: Hod's inequality}).

Thus we find that in $D$-dimensional MTBH the fundamental QNF of the massless boson and Dirac fields do not satisfy inequality (\ref{eq: Hod's inequality}). We expect that inequality (\ref{eq: Hod's inequality}) be satisfied in MTBH \cite{Hod:2006jw} owing to Hawking's temperature of the MTBH is of the same order of magnitude as the reciprocal of the characteristic length ($L$) of the spacetime. According to Hod the TTT bound (\ref{eq: Hod bound one}) is universal and we do not know the cause of its failure for the fundamental QNF of the MTBH.

\section{Effective potentials}
\label{section 4}

Following the method of Chandrasekhar's book \cite{Chandrasekhar book}, we take for the Dirac field a harmonic time dependence as in formula (\ref{eq: psi 1 2 ansatz}) to transform Eqs.\ (\ref{eq: Dirac equation topological}) into the pair of decoupled Schr\"odinger type equations
\begin{equation} \label{eq: Schrodinger equations}
 \frac{\dd^2 Z_{\pm}}{\dd \hat{r_*}^2} + \omega^2 Z_{\pm} = V_{\pm} Z_{\pm},
\end{equation} 
where 
\begin{align} \label{eq: definitions functions}
Z_{\pm} &= \textrm{e}^{i \theta/2} \bar{R}_1 \pm  \textrm{e}^{ - i \theta/2} R_2, \nonumber \\ 
\theta & = \textrm{arctan} \frac{m z}{\hat{K}}, \nonumber \\
V_{\pm} & = W^2 \pm \frac{\dd W}{ \dd \hat{r_*} } ,  \\
W & = \frac{\sqrt{z^2 -1} \left( \hat{K}^2 + (m z )^2 \right)^{3/2}}{z (\hat{K}^2 + (m z )^2) + \tfrac{\hat{K} m}{2 \omega L} z (z^2 -1) } , \nonumber  
\end{align} 
$\hat{K} =  K / L $, we define $\hat{r_*}$ by
\begin{equation}
 \frac{\dd \hat{r_*}}{\dd r_*} =  1 + \frac{z^2 -1}{2 \omega L}\frac{m \hat{K}}{\hat{K}^2 + (mz)^2} ,
\end{equation} 
and, as in Sec.\ \ref{section 2}, $r_*$ denotes the tortoise coordinate of the $D$-dimensional MTBH (see formula (\ref{eq. tortoise coordinate})).

From formulas (\ref{eq: Schrodinger equations}) and (\ref{eq: definitions functions}) we see that the effective potentials $V_{\pm}$ are complicated functions of the different parameters. Nevertheless in the massless limit we find that the formulas for $W$ and $V_{\pm}$ reduce to
\begin{align} \label{eq: potentials massless limit}
 W & =  \hat{K} \frac{\sqrt{z^2 -1 }}{z} = - \hat{K} \textrm{sech} (r_* / L), \nonumber \\
V_{\pm} & = \frac{\hat{K}^2}{\cosh^2(r_* / L)} \pm \frac{(\hat{K} / L )\sinh(r_* / L)}{ \cosh^2(r_* / L) }.
\end{align} 
Thus for the massless Dirac field the effective potentials (\ref{eq: potentials massless limit}) are of Morse type \cite{Dutt:1988susy}. In Ref.\ \cite{LopezOrtega:2007vu} it was shown that in $D$-dimensional MTBH the effective potentials of the Schr\"odinger differential equations for the massless boson fields are of P\"oschl-Teller type. We note that for many of the spacetimes for which we exactly calculate their QNF the effective potentials of Schr\"odinger type equations are of P\"oschl-Teller o Morse type (see for example Table 1 in Ref.\ \cite{Lopez-Ortega:2006my}).

\section{Discussion}
\label{section 5}

For the $D$-dimensional MTBH in Sec.\ \ref{section 2} we found that the real part of the QNF is determined by the eigenvalues of the Laplace operator (boson fields) or the Dirac operator (Dirac field) on the negative curvature manifold $\Sigma_{D-2}$. Nevertheless, up to our knowledge there are no many calculations of the spectrum of the Dirac operator on compact spin manifolds of hyperbolic type. We believe that this mathematical problem deserves detailed study. Furthermore we notice that the imaginary part of QNF (\ref{eq: QN frequencies topological}) is independent of the eigenvalues of the Dirac operator on $\Sigma_{D-2}$. This fact allows us to discuss some phenomena (see Secs.\ \ref{section 2} and \ref{section 3}) even if we do not know explicitly the value of the eigenvalues of the Dirac operator on the manifold $\Sigma_{D-2}$.

For the massless boson and Dirac fields the imaginary part of the QNF shows a different dependence on the spacetime dimension. For the boson fields the decay time depends on the spacetime dimension whereas for the Dirac field it is independent of the spacetime dimension. Also we point out that for $D \geq 6$ the massless boson fields decay faster than the massless Dirac field.

In MTBH the QNF of the Klein-Gordon, gravitational, electromagnetic, and Dirac perturbations have been calculated (see Refs.\ \cite{Birmingham:2006zx}--\cite{LopezOrtega:2007vu} and Sec.\ \ref{section 2} of this paper). Nevertheless as far as we know the QNF of the Rarita-Schwinger field have not been computed. We believe that the calculation of the QNF for this field is an interesting problem.

According to Hod the TTT bound of formula (\ref{eq: Hod bound one}) is universal  \cite{Hod:2006jw}, \cite{Hod:2007tb}--\cite{Hod:2008zz}, but we found in Sec.\ \ref{section 3} that for the fundamental QNF of the MTBH the inequality (\ref{eq: Hod's inequality}) is not satisfied (see formulas (\ref{eq: Dirac field ratio}) and (\ref{eq: bound bosons})). We believe that this puzzling result deserves detailed study.

For the $D$-dimensional MTBH, from our results and those already published we obtain that the real part of the QNF depends on the eigenvalues of the Laplace or Dirac operators on the negative curvature manifold $\Sigma_{D-2}$. These values can be different for distinct fields, also for a fixed field these eigenvalues may depend on the mode of the field. Thus the asymptotic limit of the real part of the QNF for the $D$-dimensional MTBH depend on the physical parameters of the black hole and the field (and the mode of the field). 

An interesting proposal is the so called Hod's conjecture \cite{Hod:1998vk}, it states that in the semiclassical limit the area quantum of a event horizon can be calibrated with the asymptotic value of the real part of the QNF. The facts mentioned in the previous paragraph imply that Hod's conjecture is not valid for the $D$-dimensional MTBH (as for the $D$-dimensional de Sitter spacetime \cite{LopezOrtega:2009ww}), since in this conjecture we must assume that the real part of the QNF depends only on the physical parameters of the black hole \cite{Hod:1998vk}, \cite{Maggiore:2007nq}, but it does not happen in $D$-dimensional MTBH. Thus we think that for the $D$-dimensional MTBH we must investigate whether the recent proposal of Maggiore \cite{Maggiore:2007nq} can be used to determine the area quantum of its event horizon. Work along this line is in progress.

Finally we notice that formulas (\ref{eq: QN frequencies topological}) also give the QNF of the Dirac field propagating in three-dimensional static BTZ black hole with mass $M=1$. The QNF of the Dirac field evolving in the static BTZ were previously calculated in Refs.\ \cite{Birmingham:2001pj}, \cite{Cardoso:2001hn}.

\section{Acknowledgments}

I thank Dr.\ C.\ E.\ Mora Ley and A.\ Tellez Felipe for their interest in this paper. This work was supported by CONACYT M\'exico, SNI M\'exico, EDI-IPN, COFAA-IPN, and Research Projects SIP-20090952 and SIP-20091344.

\end{document}